# Dual electro-optic frequency comb photonic thermometry

ADAM J. FLEISHER,[1,*] ZEESHAN AHMED,[2] TOBIAS HERMAN,[2] AND MATTHEW R. HARTINGS[3]

[1]Material Measurement Laboratory, National Institute of Standards and Technology, 100 Bureau Drive, Gaithersburg, MD 20899, USA
[2]Physical Measurement Laboratory, National Institute of Standards and Technology, 100 Bureau Drive, Gaithersburg, MD 20899, USA
[3]Department of Chemistry, American University, 4400 Massachusetts Avenue NW, Washington, DC 20016, USA
*Corresponding author: adam.fleisher@nist.gov

We report a precision realization of photonic thermometry using dual-comb spectroscopy to interrogate a π-phase-shifted fiber Bragg grating. We achieve read-out stability of 7.5 mK at 1 s and resolve temperature changes of similar magnitude—sufficient for most industrial applications. Our dual-comb approach enables rapid sensing of dynamic temperature, and our scalable and reconfigurable electro-optic generation scheme enables a broad sensing range without laser tuning. Reproducibility on the International Temperature Scale of 1990 is tested, and ultimately limited by the frequency reference and check-thermometer stability. Our demonstration opens the door for a universal interrogator deployable to multiple photonic devices in parallel. Applications include on-chip measurements to simultaneously evaluate quantities like temperature, pressure, humidity, magnetic field and radiation dose.

The past two decades have witnessed tremendous growth in the field of photonic devices [1–4]. In turn, these devices now enable new sensing motifs—from bio-sensing to gravity and magnetic field sensing to thermodynamic metrology. However, as the underlying technology has matured, sensor interrogation has emerged as a key bottleneck to wider adoption [1]. To date, optical approaches to photonic sensor read-out have been forced to choose between one of two distinct capabilities: high precision or broad coverage. Precision measurements, like those performed in metrology laboratories, often rely on complex techniques for tight laser locking [5–7] while industrial sensing, which demands broader coverage, prefers user-friendly solutions like swept-wavelength laser spectroscopy [4,8–10]. These examples of disparate applications highlight that the choice of interrogation technique takes on an outsized importance when designing functional photonic sensors.

To bridge the complexity gap between tight laser locking and swept-wavelength laser scanning, new approaches to photonic sensor read-out have emerged. Optical self-heterodyne [11] or interleaved [12] read-out of a chirped-pulse laser source has been applied to accelerometry [13]. Also, dual-comb spectroscopy [14] is a powerful tool for myriad physical sensing (e.g., static strain [15], distributed sensing [16]). Recently, dual-comb spectroscopy was applied to photonic thermometry using two mode-locked lasers and an adaptive phase-correction scheme to interrogate a πFBG [17]. Currently underexplored is the application of dual electro-optic frequency comb spectroscopy [18] as a flexible and potentially universal interrogator for photonic devices. Compared to chirped-pulse waveform generation, electro-optic combs generated by phase and/or intensity modulation offer comparably fast µs or ns read-out times [19–22] with superior optical bandwidth, scalability, frequency conversion and on-chip integration [23–26] while requiring only a single, RF drive signal and correspondingly lower-bandwidth modulators. Compared to mode-locked lasers, electro-optic combs offer improved flexibility and frequency agility without sacrificing the potential for very broad optical bandwidth.

Here we explore a general solution for photonic thermometry, working towards a universal interrogator with the following performance metrics [8,9]: Sensitivity of several GHz K$^{-1}$ over an industrially relevant range of 80 K to 1200 K, low read-out uncertainty of ≤10 mK, high time resolution of ≤1 µs, read-out stability for >1 s and a low size, weight, power and cost footprint. We use all-fiber dual-comb spectroscopy to probe a narrow-band πFBG resonance with high spectral and temporal resolution and without laser tuning or moving parts. We generate power-leveled electro-optic frequency combs through a combination of intensity and phase modulation, beginning from a continuous-wave seed laser. The result is a highly scalable and reconfigurable interrogator for rapid sensing with high Q-factor photonic devices. Specifically, we demonstrate temperature measurements near 303.15 K (30 °C) and 273.15 K (0 °C), respectively, with GHz K$^{-1}$ sensitivity, sub-10 mK stability at 1 s, and high precision over a 7 K wide window. We also show dynamic temperature sensing with time resolution <1 s.

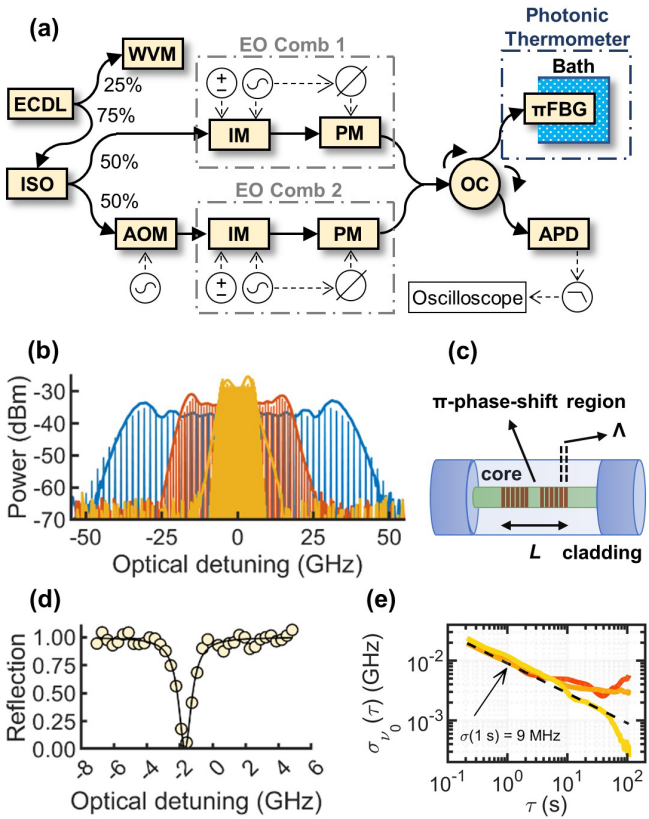

**Fig. 1. Dual-comb photonic thermometry.** (a) Experimental set-up for dual electro-optic (EO) comb generation and photonic thermometry. Abbreviations: ECDL, external cavity diode laser; WVM, wavelength meter; ISO, optical isolator; AOM, acousto-optic modulator; IM, intensity modulator; PM, phase modulator; OC, optical circulator; πFBG, π-phase-shifted fiber Bragg grating; APD, avalanche photodiode. (b) Optical spectrum analyzer traces for EO combs of various repetition rates (bold lines) compared to simulations (fine sticks). Blue, $f_{rep}$ = 1.92 GHz; red, $f_{rep}$ = 0.92 GHz; green, $f_{rep}$ = 0.32 GHz. (c) Schematic and cross-section view of a πFBG with period Λ and length $L$ [27]. (d) Normalized dual-comb reflection spectrum for the πFBG immersed in the gallium (Ga) melting-point cell at $T_{90}$ = 302.9146 K. (e) Allan deviation (σ) for the fitted πFBG center frequency ($ν_0$) vs. laboratory time (τ) at the Ga melting point. Three trials (solid lines) scatter about the $\sqrt{\tau}$ model (dashed line) at τ > 10 s, attributed to the limited number of acquisitions per trial.

The set-up for dual-comb photonic thermometry is illustrated in **Fig. 1**(a). The continuous-wave output from an external-cavity diode laser (ECDL) operating at a wavelength near 1550 nm was split in fiber, with 25 % going to a wavelength meter (WVM) with ±15 MHz absolute accuracy and 75 % going to seed dual-comb generation. The laser seed was split to generate two electro-optic (EO) combs, each using one intensity modulator (IM) and one phase modulator (PM). The IMs act to carve a portion of the phase chirp imparted by the PM driven at high modulation depth of $β_m = |\phi_0/\pi|$ = 5.25 (where $\phi_0$ is the phase shift and PM $V_\pi$ = 3.8 V at an RF drive frequency of $f$ = 1 GHz). To shift the dual-comb interferogram away from baseband in the down-converted RF domain, an acousto-optic (AO) modulator is added to one of the electro-optic combs and driven at an offset frequency of $\Delta f_0$ = +55 MHz. All RF signals used to generate the combs were referenced to a single 10-MHz clock signal.

After recombination in fiber, the dual combs were transmitted to port-1 of an optical circulator (OC) and sent to the photonic thermometer connected at port-2—a πFBG immersed in a bath under test. Reflection from the photonic thermometer was collected at port-3 and detected by an avalanche photodiode (APD). The APD output was low-pass filtered and dual-comb interferograms were measured on a fast oscilloscope at a sampling rate of $8 \times 10^8$ s$^{-1}$. For spectral normalization of the reflection signal, the OC was bypassed and a reference dual-comb interferogram recorded.

Power spectra of example EO combs generated by our two-modulator design are shown in **Fig. 1**(b), recorded on an optical spectrum analyzer with 4 GHz resolution. Corresponding simulations show fine sticks which represent the individual comb modes at higher resolution. Generally, our EO combs comprise 40 comb teeth within a 5 dB power-leveled range. Here we show that this modest comb generation is sufficient to interrogate a high $Q$-factor photonic device—and note that our chosen approach is highly scalable and relevant for measurement applications where both precision and breadth are required. As definitive examples of EO comb scalability, others have generated self-reference optical frequency combs beginning from similar pulse carving modulation schemes (e.g., [28,29]).

Here we use EO combs to interrogate a photonic device for precision temperature sensing. During measurements, the πFBG (**Fig. 1**(c)) was immersed in a bath-under-test like the Gallium (Ga) melting-point cell. A normalized dual-comb intensity reflection spectrum recorded with the πFBG immersed in the Ga melting-point cell is shown in **Fig. 1**(d). Each data point is calculated from the intensity of a unique dual-comb tooth pair, and the optical detuning axis is reconstructed from the programed RF modulation frequencies that drive the EO comb modulators (i.e., the comb repetition rates, $f_{rep}$, and offset frequency, $\Delta f_0$). Here, EO comb 1 had a repetition rate of $f_{rep,1}$ = 320.183 MHz, and the difference in repetition rate between combs was $\Delta f_{rep} = f_{rep,2} - f_{rep,1}$ = +1 MHz.

In **Fig. 1**(d), we fit the reflection spectrum recorded at 1 ms of dual-comb interrogation time using a Lorentzian function for the πFBG feature and a quadratic function for any residual baseline. The fitted half-width at half-maximum for the πFBG resonance is $\Gamma_0$ = 440 MHz, yielding an optical quality factor of $Q$ = 22 000 at 1550 nm. Such a high $Q$-factor photonic thermometer underscores the need for flexible electro-optic dual-comb generation using programmable values of $f_{rep}$ < 1 GHz.

As one of the defining points on the International Temperature Scale of 1990 (ITS-90), the Ga melting point cell delivers a stable temperature of $T_{90}$ = 302.9146 K ($T_{90}$ = 29.7646 °C) [30]. Here we use the Ga melting point to evaluate stability of the dual comb system for temperature read-out. Shown in **Fig. 1**(e) are a series of Allan deviation measurements for the fitted center frequency ($ν_0$) of the πFBG reflection feature (e.g., **Fig. 1**(d)). The series shows averaging behavior consistent with the $\sqrt{\tau}$ model for at least 10 s of laboratory time (τ), before the Allan deviation traces scatter. From the data in **Fig. 1**(e), we report stability for $ν_0$ at 1 s to be σ(1 s) = 9 MHz which corresponds to a temperature read-out stability at 1 s of 7.5 mK for the πFBG (see **Fig. 2** for conversion factor, β). In silicon devices with a larger thermo-optical coefficient, this would translate into read-out stability of ≈750 μK [1,8]. The relative stability of the dual-comb interferometer is likely a source of drift on the time scale of **Fig. 1**(e), and real-time schemes that utilize phase-locked loops or software-based corrections to push through this limit are known [21,31,32]. Deployment of these interventions within the system described would immediately enhance long-term stability.

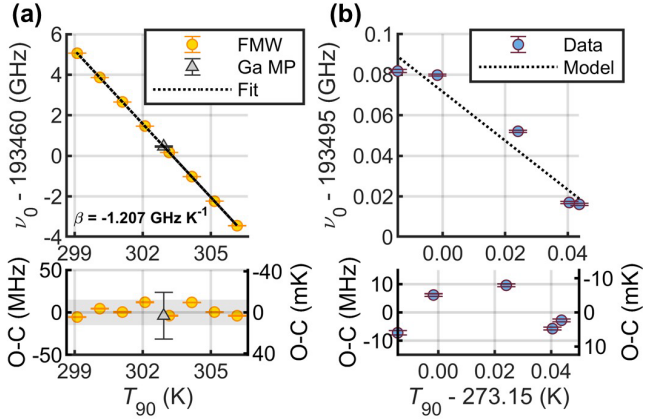

**Fig. 2. Precision photonic thermometry.** (a) Measured πFBG $\nu_0$ as a function of field metrology well temperature (FMW, gold circles) established with reference to the Ga melting point. A linear fit (top; black dotted line) yields a πFBG temperature-tuning coefficient of $\beta = -1.207$ GHz K$^{-1}$. Scatter in the fitted residuals (bottom; O–C, observed-minus-calculated) is consistent with the accuracy of the wavelength meter (bottom; gray rectangle, ±15 MHz). The larger error bar for the Ga melting-point data (Ga MP, gray triangles) illustrates experimental reproducibility. (b) mK precision for small changes near $T = 273.15$ K. The model (top; black dotted line) slope is equal to the fitted value from (a). Scatter in the residuals (bottom; O–C) is consistent with the field metrology well stability (±10 mK) and calibrated platinum resistance thermometer (PRT) accuracy (10 mK).

The optical bandwidth of our EO combs enables πFBG $\nu_0$ read-out over several K without the need for laser tuning or demanding laser stabilization at each temperature step. Plotted in **Fig. 2**(a) are the measured $\nu_0$ fitted when the πFBG temperature was tuned from $T_{90} \approx 299$ K to $T_{90} \approx 306$ K using a programmable field metrology well (FMW, gold circles). Each value of $\nu_0$ is the average of 1000 measurements, recorded in a laboratory time of ≈200 s per point. The temperature scale was linked to ITS-90 through repeated πFBG measurements in the Ga melting point cell (Ga MP, gray triangle) and subsequent calibration of the co-located platinum resistance thermometer (PRT, or check thermometer). The πFBG thermal tuning coefficient was fitted on $T_{90}$ to a value of $\beta = -1.207$ GHz K$^{-1}$.

The residuals from the linear fit plotted in the bottom of **Fig. 2**(a) provide a further estimate of precision for our prototype dual EO comb photonic thermometer. Without read-out calibration, we show that $\nu_0$ for the πFBG resonance is reproducibly retrieved and consistent with the linear trend, limited by the absolute uncertainty of the wavelength meter (±15 MHz, gray shaded region). The standard deviation of the residuals is also 15 MHz (or 0.12 pm), which, using our fitted value for β, yields an estimated $T_{90}$-uncertainty of 13 mK. The larger error bar plotted for the Ga melting-point data (gray triangle) is the standard deviation of four repeated measurements recorded over several days, and illustrates experimental reproducibility (whereas error bars plotted for the field metrology well data show standard deviation.) Importantly, the wavelength-meter-limit can be readily overcome by linking the continuous-wave laser frequency to a precision optical reference like a stabilized optical cavity, molecular absorption feature, or self-referenced optical frequency comb.

In **Fig. 2**(b), we test the instrument precision for small (mK) changes in temperature, again using the field metrology well and a calibrated PRT. Plotted in blue-filled red circles are the fitted values of $\nu_0$ as a function of PRT temperatures near 273.15 K. When the data points shown in the top of **Fig. 2**(b) were compared to the linear model with $\beta = -1.207$ GHz K$^{-1}$, the standard deviation of the residuals shown in the bottom of **Fig. 2**(b) yielded an estimated precision of 7.4 MHz. The scatter is similar to the stability at $\tau > 100$ s estimated from the Allan deviation series in **Fig. 1**(e) to be ≤3 MHz. Further, the scatter is likely influenced by the calibrated PRT accuracy of 10 mK and may reflect the intrinsic limitations of the chosen check thermometer. Other sources of error may include instabilities in the field metrology well near the 273.15 K set point arising from its non-ideal environment—high relative humidity and an ambient laboratory temperature near 300 K. Indeed, the manufacturer-specified stability of the field metrology well is also ±10 mK over the full range—suggesting that our measurements are limited by a combination of equipment often used in metrology laboratories. Overall, the stability (**Fig. 1**(e)), range (**Fig. 2**(a)), and precision (**Fig. 2**(b)) tests of our dual frequency comb photonic thermometry demonstration are consistent.

Data in **Fig. 3** illustrate sub-s time resolution for our dual-comb, πFBG sensor. **Fig. 3**(a) shows a waterfall plot of the fitted πFBG reflection spectrum near resonance. As the measurement proceeds over 200 s, the πFBG resonance moves to lower optical detuning (i.e., a lower absolute optical frequency). The values of $\nu_0$ retrieved from the models are plotted as a light blue trace in the top of **Fig. 3**(b), using the tick marks on the left y-axis as indicated by the light blue arrow. Also plotted in the gray trace is the programmed field metrology well temperature ramp, connected with the right y-axis as indicated by the gray arrow. As the temperature at the πFBG is increased from $T \approx 277.65$ K to $T \approx 278.15$ K ($T \approx 4.5$ °C to $T \approx 5.0$ °C), we readily resolve the πFBG resonance shifting to lower optical frequencies. For the nearly 1000 data points plotted in the light blue curve in **Fig. 3**(b), the integration time per point was 1 ms (1000 interferograms, $\Delta f_{\text{rep}} = 1$ MHz) and the point-by-point acquisition rate was 4.7 Hz (i.e., the 200 s measurement window is 99.5 % dead time). Dead time can be eliminated by improved acquisition software and hardware, triggering and data storage (e.g. [21,31,32]).

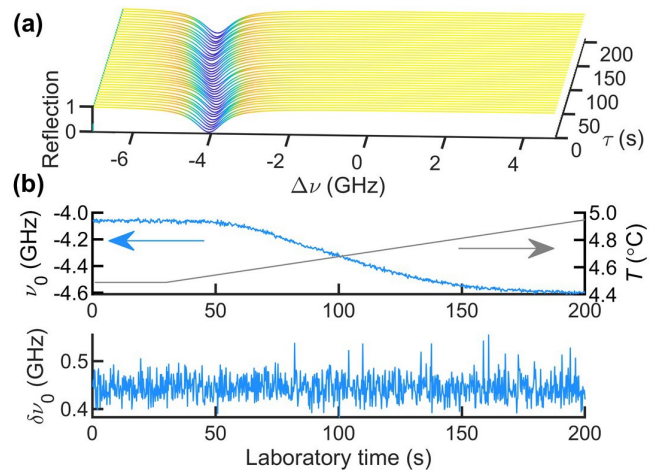

**Fig. 3. Dynamic photonic thermometry.** (a) Waterfall plot of the fitted models for each dual-comb πFBG reflection spectrum vs. τ. (b) Time-resolved read-out of $\nu_0$ (top, left y-axis; light blue) and programmed field metrology well ramp (top, right y-axis; gray). Measured half-width at half-maximum ($\Gamma_0$) for the πFBG resonance vs. τ (bottom; light blue).

**Figure 3**(b) also shows the fitted πFBG resonance half-width at half-maximum ($\Gamma_0$) during the temperature ramp. It is constant as a function of τ at a value of $\Gamma_0$ = 440 MHz ± 20 MHz, meaning that the $Q$-factor is virtually unaffected by the dynamic temperature.

As a read-out technique, dual-comb spectroscopy has known trade-offs between the achievable optical bandwidth (Δν), comb tooth spacing (repetition rate, $f_{rep}$) and acquisition rate (Δ$f_{rep}$) [14]. For sensing, reconfigurable EO comb generation enables frequency-agile exploration of these trade-offs—significantly more so than with mode-locked lasers. Our dual-comb interrogator covered the temperature range of 7 K with $M$ = 40 teeth generated using two modulators per comb. At $f_{rep,1}$ = 320.183 MHz, Δν = $Mf_{rep,1}$ = 12.8 GHz. Therefore, our highest acquisition rate is Δ$f_{rep}$ ≤ $(f_{rep,1})^2/(2\Delta\nu)$ = 4 MHz. This translates to a minimum time resolution of 250 ns.

Broad deployment of photonic thermometers requires an interrogation system that can support GHz K$^{-1}$ sensitivity over a broad temperature range (80 K to 1200 K), with high accuracy (≈10 mK), time resolution (≤1 μs), and long-term stability (>1 s). We report here that dual EO comb spectroscopy provides a near ideal mix of measurement accuracy, time resolution, information content and scalability. In the application presented, we describe a system for interrogating a photonic device with a $Q$-factor of 220 000 over a 7 K temperature range with a readout stability of 7.5 mK at 1 s and a integration time of 1 ms. This performance shows that our applied dual EO comb read-out can meet the technical specifications for most industrial and precision calibration applications. Additionally, as our precision is limited by equipment that are standard in photonic thermometry research labs, we show that dual EO comb spectroscopy is immediately relevant for precision temperature metrology. The $T_{90}$ uncertainty observed in this study is ±13 mK and is due, in large part, to the uncertainty of the wavelength meter. This limitation can be overcome via a link to a precision optical reference.

Electro-optic combs can be generated on-chip [25,26], and can provide access to information-rich spectra that can be used to extract the real and imaginary components of the refractive index profile—depending on the dual-comb interferometer configuration [14]. Combined with physics-based models of photonic sensors this could be utilized to build advanced machine learning models such as deep neural networks or hidden Markov models that incorporate bandgap physics, material chemistry knowledge and coupled mode theory to create self-calibrating algorithms to compensate for long-term drift in the baseline refractive by incorporating information extracted from the spectral signatures [33]. Importantly, these models can then be applied in a manner such that read-outs from sensor networks can be rationally interpreted in systems where traceability and transparency are of primary concern [1].

**Funding.** National Institute of Standards and Technology.

**Acknowledgments.** The authors acknowledge J. T. Hodges, D. M. Bailey and R. P. Fitzgerald of NIST for commenting on the paper.

## References

1. S. Dedyulin, Z. Ahmed, and G. Machin, Meas. Sci. Technol. **33,** 092001 (2022).
2. J. Wang and Y. Long, Sci. Bull. **63,** 1267 (2018).
3. C. Li, D. Liu, and D. Dai, Nanophotonics **8,** 227 (2019).
4. V. Sudhir, R. Schilling, S. A. Fedorov, H. Schütz, D. J. Wilson, and T. J. Kippenberg, Phys. Rev. X **7,** 031055 (2017).
5. G. Gagliardi, M. Salza, S. Avino, P. Ferraro, and P. De Natale, Science **330,** 1081 (2010).
6. J. Ricker, K. O. Douglass, J. Hendricks, S. White, and S. Syssoev, Meas. Sensors **18,** 100286 (2021).
7. A. Reihani, E. Meyhofer, and P. Reddy, Nat. Photon. **16,** 422 (2022).
8. H. Xu, M. Hafezi, J. Fan, J. M. Taylor, G. F. Strouse, and Z. Ahmed, Opt. Express **22,** 3098 (2014).
9. T. Briant, S. Krenek, A. Cupertino, F. Loubar, R. Braive, L. Weituschat, D. Ramos, M. J. Martin, P. A. Postigo, A. Casas, R. Eisermann, D. Schmid, S. Tabandeh, O. Hahtela, S. Pourjamal, O. Kozlova, S. Kroker, W. Dickmann, L. Zimmermann, G. Winzer, T. Martel, P. G. Steeneken, R. A. Norte, and S. Briaudeau, Optics **3,** 159 (2022).
10. E. Shafir, G. Berkovic, A. Fedotov-Gefen, A. Ravid, S. Zilberman, and Y. Schweitzer, in 2017 Conference on Lasers and Electro-Optics Pacific Rim (Optica Publishing Group, 2017), paper s1150.
11. D. A. Long, A. J. Fleisher, D. F. Plusquellic, and J. T. Hodges, Phys. Rev. A **94,** 061801 (2016).
12. D. A. Long, B. J. Reschovsky, T. W. LeBrun, J. J. Gorman, J. T. Hodges, D. F. Plusquellic, and J. R. Stroud, Opt. Lett. **47,** 4323 (2022).
13. F. Zhou, F. Zhou, Y. Bao, Y. Bao, R. Madugani, D. A. Long, J. J. Gorman, J. J. Gorman, and T. W. LeBrun, Optica **8,** 350 (2021).
14. I. Coddington, N. Newbury, and W. Swann, Optica **3,** 414 (2016).
15. R. Zhang, Z. Zhu, G. Wu, and G. Wu, Opt. Express **27,** 34269 (2019).
16. X. Zhao, J. Yang, J. Liu, H. Shao, X. Zhang, Q. Li, and Z. Zheng, IEEE Trans. Instrum. Meas. **69,** 5821 (2020).
17. Y. Guo, M. Yan, Q. Hao, K. Yang, X. Shen, and H. Zeng, Front. Inf. Technol. Electron. Eng. **20,** 674 (2019).
18. D. A. Long, A. J. Fleisher, K. O. Douglass, S. E. Maxwell, K. Bielska, J. T. Hodges, and D. F. Plusquellic, Opt. Lett. **39,** 2688–2690 (2014).
19. N. B. Hébert, V. Michaud-Belleau, C. Perrella, G.-W. Truong, J. D. Anstie, T. M. Stance, J. Genest, and A. N. Luiten, Phys. Rev. Appl. **6,** 044012 (2016).
20. E. L. Teleanu, V. Durán, and V. Torres-Company, Opt. Express **25,** 16427 (2017).
21. P. Guay, J. Genest, and A. J. Fleisher, Opt. Lett. **43,** 1407–1410 (2018).
22. P.-L. Luo and I.-Y. Chen, Anal. Chem. **94,** 5752 (2022).
23. V. Torres-Company and A. M. Weiner, Laser Photon. Rev. **8,** 368 (2014).
24. A. Parriaux, K. Hammani, and G. Millot, Adv. Opt. Photon. **12,** 223 (2020).
25. A. Shams-Ansari, M. Yu, Z. Chen, C. Reimer, M. Zhang, N. Picqué, and M. Lončar, Commun. Phys. **5,** 88 (2022).
26. M. Yu, D. Barton III, R. Cheng, C. Reimer, P. Kharel, L. He, L. Shao, D. Zhu, Y. Hu, H. R. Grant, L. Johansson, Y. Okawachi, A. L. Gaeta, M. Zhang, and M. Lončar, Nature (2022), https://doi.org/10.1038/s41586-022-05345-1.
27. S. Deepa and B. Das, Sens. Actuators A Phys. **315,** 112215 (2020).
28. K. Beha, D. C. Cole, P. Del'Haye, A. Coillet, S. A. Diddams, and S. B. Papp, Optica **4,** 406 (2017).
29. E. Obrzud, M. Rainer, A. Harutyunyan, B. Chazelas, A. Ghedina, E. Molinari, S. Kundermann, S. Lecomte, F. Pepe, F. Wildi, F. Bouchy, and T. Herr, Opt. Express **26,** 34830 (2018).
30. J. V. Pearce, P. P. M. Steur, W. Joung, F. Sparasci, G. Strouse, J. Tamba, and M. Kalemci, "Guide to the realization of the ITS-90, metal fixed points for contact thermometry," (2018), Bureau International des Poids et Mesures (BIPM), Consultative Committee for Thermometry (CCT) under the auspices of the International Committee for Weights and Measures (CIPM), last updated 20 October 2021.
31. A. J. Fleisher, D. A. Long, Z. D. Reed, J. T. Hodges, and D. F. Plusquellic, Opt. Express **24,** 10424 (2016).
32. K. Fdil, V. Michaud-Belleau, N. B. Hébert, P. Guay, A. J. Fleisher, J.-D. Deschênes, and J. Genest, Opt. Lett. **44,** 4415 (2019).
33. Z. Ahmed, Sens. Actuators A Phys. **347,** 113872 (2022).